# Effect of thermal treatments in high purity Ar on the oxidation and tribological behavior of arc-PVD c-Al$_{0.66}$Ti$_{0.33}$N coatings

G.C. Mondragón Rodríguez*, Y. Chipatecua-Godoy[1], N. Camacho, D.G. Espinosa-Arbeláez, J.M. González Carmona, J. González-Hernández, J.M. Alvarado-Orozco*

CONACyT, Center of Engineering and Industrial Development, CIDESI, Surface Engineering Department, Querétaro, Av. Pie de la Cuesta 702, 76125 Santiago de Querétaro, Mexico.

[1]Centro de Investigación y Estudios Avanzados del IPN, Lib. Norponiente 2000, Fracc. Real de Juriquilla, 76230 Querétaro, Qro. Mexico.

*Corresponding author(s), e-mail: guillermo.mondragon@cidesi.edu.mx, juan.alvarado@cidesi.edu.mx Tel: +52 (442) 2119800 ext. (5313).

## Abstract

The effect of temperature in high purity Ar (low oxygen partial pressure) on the oxidation and crystal phase evolution of c-Al$_{0.66}$Ti$_{0.33}$N arc-PVD coatings was investigated. High temperature tribology in Ar jet and adhesion behavior of the oxidized coating were addressed. The use of Ar protects the coating from the oxidation reactions and allowed to shed light into the details and paths of the nitride oxidation process. The c-Al$_{0.66}$Ti$_{0.33}$N nitrides were slightly oxidized in Ar even at high temperature (700 to 900 °C). The surface chemistry evaluation shows very thin Al- and Ti-based oxide layers formation after treatment at 700 °C in Ar. However, the onset formation of the rutile TiO$_2$ oxide was detected only after treatment at 900 °C and was clearly found by low-angle XRD diffraction after 1000 °C. The XPS surface analysis of the oxidized samples at 800°C indicated the formation of extremely thin layers mainly composed of mixed oxides (α-Al$_2$O$_3$ γ-Al$_2$O$_3$ and rutile TiO$_2$). The gradual changes in the chemical composition of the oxidized coatings observed at 900 °C clearly demonstrate the formation of gradient boundaries between Al-rich and Ti-rich oxy-nitride layers. After exposure at 1000 °C at least four oxides and oxi-nitrides layers and interlayers were found. Additionally, it is suspected that the transition oxide phases such as c-TiO, c-Al$_{0.54}$Ti$_{2.46}$O$_{0.28}$N$_{4.58}$, α-Al$_2$TiO$_5$ and TiO$_x$N$_y$ might be precursors leading to the formation of thermodynamically stable rutile and alumina phases. Surface inspection after high temperature tribology in Ar jet showed formation of tensile cracks in the wear tracks and wear processes due to high size particle adhesion-abrasion and low size particle micro-plowing of hard nitrides, oxi-nitrides and brittle oxide particles. Adhesion tests performed after high temperature tribology in Ar showed critical load decrease related to cohesive and adhesive damages at the contact point of the scratch track.

1. Introduction

The mechanical strength and oxidation resistance of c-$Ti_{(1-x)}Al_xN$ coatings relies upon its chemistry and microstructure. In c-$Ti_{(1-x)}Al_xN$ nitrides, the Al-content can vary between 0.25 and ~ 0.92 [1]. However, only nitrides with a maximum Al content of $x = 0.6$ can be assumed to be stable in a cubic structure. Slightly higher Al-content (> 0.7) can lead to the formation of h-AlN, negatively impacting the coating hardness [2]. Hardness improvement can be induced either by spinodal decomposition or grain refinement. Spinodal decomposition depends on nitride splitting into its single cubic nitrides [3-6]. This decomposition process is strongly conditional on the coating composition and the temperature of exposure [7, 8], *e.g.* c-$Ti_{0.36}Al_{0.64}N$ starts to decompose at 850 °C [9]. On the other hand, grain refinement produced by Al-concentrations up to 0.6 has a favorable effect on hardness [9]. This improvement is mainly caused by the Hall-Petch effect [11-13]. Other studies explain hardness decrease (or increase) as a combination of microstructural characteristics such as grain size effect, texture and residual stresses related to the coating deposition parameters [14].

Oxidation resistance of the ternary nitride is increased by Al-addition into the TiN [15]; either small amounts of Al ($Ti_{0.6}Al_{0.3}N$) or high Al-contents ($Ti_{0.4}Al_{0.6}N$) have a positive impact on the oxidation resistance [16]. The improved oxidation performance is due to the formation of a protective layer (*e.g.* $Al_2O_3$) upon exposure to temperature and oxygen. Following the reaction, a dense and homogenous oxide layer forms with time, if enough Al supply is available, further retarding ion diffusion and diminishing coating degradation. As a consequence of hardness improvement and better oxidation resistance, huge impacts on the performance of coated parts, components and tooling has been accomplished. For instance, TiAlN coated drills achieved 269 % more holes compared to the work reached with TiN coated tools [15]. Similar performance achievements have been reported elsewhere for machining processes such as turning [17] and milling [18].

Thermal treatments under controlled low oxygen partial pressures (e.g., vacuum conditions), temperature and time have been applied as an annealing step to investigate the age-hardening processes that might take place during coating service [9, 10]. Thermal treatments inducing hardness enhancement start at 500 °C [10], but significant hardness improvement is only reached after exposure to temperatures between 700 and 1000 °C [9, 10, 19, 20]. Treatments at temperatures higher than 1000 °C cause a decrease in hardness, which has been associated with the formation and growth of incoherent w-AlN [9, 10]. Particularly, nitrides with high Al contents ($Ti_{0.33}Al_{0.66}N$) displayed peak hardness condition after thermal treatments at 900 °C in comparison with nitrides containing less Al [20]. Johansson *et al* [20] also described the age-hardening effects on $Ti_{(1-x)}Al_xN$ system (for x = 0.31, 0.47, 0.67) caused by the harsh temperature and pressure conditions during a turning operation. In practice, conventional coated metal tools with Al-containing nitrides are exposed to ambient air and elevated temperatures thus providing the best conditions for oxidation reactions. Degradation studies of c-$Ti_{(1-x)}Al_xN$ coatings in such highly oxidizing environments have been carried out in the past; for instance, arc-PVD $Ti_{0.5}Al_{0.5}N$ coating has been severely oxidized after exposure to 850 °C/10h [21]. Pfeiler *et al* reported full oxidation of $Ti_{0.33}Al_{0.67}N$ after annealing at 900 °C/5h in air [22]. As an alternative to protect $c-Ti_{(1-x)}Al_xN$ coatings from oxidation and degradation, machining operations could be carried out in the presence of inert gases such as $N_2$, $CO_2$, Ar or any mixture of these. The term gas-cooled machining has been coined for environmental friendly machining where a gas lubricant is used for machining instead of solid or liquid lubricant [23]. In addition to the positive environmental impact, gas cooled machining was proposed in the 30s for improving machinability and lubricity at the tool-workpiece-chip interface. Recently, different attempts to improve high speed dry machinability of titanium alloys using uncoated carbides in Ar-rich environments have been carried out in order to avoid the chemical reactions at the tool - chip and tool – workpiece [24]. It has been reported that Ar promoted wear at the cutting edge negatively impacting the performance of the tool and decreasing operative tool life, which was mainly due to poor heat transfer. In other cases, the use of argon has been successfully applied as lubri-cooling protection of ceramic tools for dry machining of Inconel 715 [25]. To our knowledge ,inert atmospheres have not been proposed as medium to protect $c-Ti_xAl_{1-x}N$ based coatings from oxidation. From this perspective, it would be relevant to evaluate the degree of oxidation under such working conditions. For these reasons, this research was focused on investigating the crystal phase evolution during oxidation treatments of the arc-PVD $c-Al_{0.66}Ti_{0.33}N$ coatings treated in high purity Ar at temperatures up to 1000 °C. Additionally, wear behavior and crystal phase transformation upon temperature of $c-Al_{0.66}Ti_{0.33}N$ under argon jet was studied.

## 2. Experimental

*2.1 Coating deposition*

c-Al$_{0.66}$Ti$_{0.33}$N hard coatings were deposited on 10 x 20 x 1 mm-alumina substrates and 1 in-mirror polished AISI M2 steel using an arc-PVD coater from Oerlikon model Domino Mini. Alloy targets (100 mm Ø) of Al/Ti:0.66/0.33 at. % and 99.5 % purity from Plansee were used. The alumina substrates were ultrasonically cleaned, dried and mounted in a three-fold rotation system. Firstly, the substrates were coated with a ~ 1µm thick Ti layer, hereafter called Ti-interlayer, to improve the adhesion of the nitride. Then, the Ti-metalized substrates were heated up to about 330 °C using a 9 KW heating system in the vacuum chamber. Substrates temperature was measured by thermocouples located near the planetary. Before the coating step, the substrates surface were cleaned via erosion with Ar+ ions using the AEGD (Arc Enhanced Glow Discharge) – Etching process. The deposition time was maintained for about 1 h applying 80 V bias. Upon coating, the working temperature reached ~ 430 °C.

*2.2 Isothermal treatment in Ar*

The isothermal treatments at 700, 800, 900 and 1000 °C were carried out in a thermogravimetric analyzer (TGA) system using high purity argon (99.997 %). The isothermal TGA experiments were conducted in a Setsys equipment from Setaram. The furnace chamber of the TGA was evacuated for 20 min and then filled with 200 ml/min of Ar. Subsequently, the Ar flow was changed to 20 ml/min for 10 min and, finally, heated at a ramp of 50 °C/min in Ar up to the selected treatment temperature. The TGA data was separately collected during the heating ramp and at each isothermal temperature for 5h. The obtained TGA results were normalized to the calculated geometric surface area of the coated Al$_2$O$_3$ samples (approx. 4.53 cm$^2$).

*2.3 Coating analysis*

Roughness of the c-Al$_{0.66}$Ti$_{0.33}$N-Al$_2$O$_3$ and c-Al$_{0.66}$Ti$_{0.33}$N-AISI M2 steel systems was measured by means of contact profilometry using a DektaktXT equipment from Bruker. Roughness measurements were carried out using a Stylus tip with 12.5 µm radius over 2 mm distance and repeating the process 5 times. The crystal phase measurements prior and subsequent to the oxidation treatments were carried out in a SmartLab XRD diffractometer from Rigaku in PB/PSA mode Parallel Beam/Parallel Slit Analyzer with a grazing incidence angle ($\omega = 0.2°$) 2θ geometry and Cu Kα radiation (40 kV, 44 Amp) at a step size of 0.02°. The diffraction patterns were indexed using the Integrated X-ray powder diffraction Software PDXL from Rigaku.

The X-ray Photoelectron Spectroscopy (XPS) analysis of the as-coated and oxidized samples was assessed in an instrument from Physical Electronics model PHI 5000 VersaProbe II using monochromatic Al-Kα radiation (1486.6 eV). The spot size, the power and the X-Ray voltage were 100 µm, 25 W and 15 V, respectively. The depth profiling of the oxidized films was performed using 2 kV $Ar^+$ ions over a 3x3 mm sputtering area and 1min erosion time. The surface erosion was ~ 6.25 nm/min leading to ~ 500 nm coating depth analysis. Binding energy values were corrected with reference to the C1s peak at 284.6 eV. The Ti spectra were fitted with Ti$2p_{3/2}$ and Ti $2p_{1/2}$ branches spin-orbit splitting of 5.55 eV. The peak fitting analysis was performed using the software AAnalyzer®.

The initial state of the coating and the sample treated at 1000 °C were analyzed in a TEM JEM 2200 FS+CS from JEOL applying 200 kV and an EDX Inca analyzer from Oxford. For this purpose, FIB-cuts of the cross-sections were previously prepared. A thin Au thin layer was deposited on top of the c-$Al_{0.66}Ti_{0.33}N$ coating and then the FIB-cut was carried out using $Ga^+$ ions.

*2.4 Tribological study*

Pin-on-disk wear experiments were carried out applying an argon jet (99.997 % purity) at RT, 700, 800, & 900 °C. A commercial High Temperature Tribometer from Anton Paar, model THT 1000 with adapted Ar jet to the Instrument to protect the coating surface during tribology was used. Prior to these experiments arc-PVD c-Al$_{0.66}$Ti$_{0.33}$N coatings were deposited, following the deposition process mentioned in section 2.1, on mirror polished AISI M2 steel without any interlayer. Alumina balls of 6-mm diameter were used and a normal load of 5 N was applied at a linear speed of 7.37 cm/s resulting in a 3.5-mm wear track radius. Samples were heated at the selected temperature in Ar prior to the pin-on-disk test. Friction coefficient against distance and cycle number were recorded. Argon jet was applied during heating, the pin-on-disk experiments, and cooling after 10,000-cycles test. Arc PVD coating-AISI M2 systems were analyzed after tribological studies via grazing incidence XRD and optical imaging. Additionally, the Al$_{0.66}$Ti$_{0.33}$N coating samples were scratch tested after Ar-tribology in a Revetest equipment model RST$^3$ from Anton Paar. Scratch tests were carried out using a Rockwell C (200 µm) indenter applying 1 – 35 N in 3.5 mm at 7 mm/min load gradient.

## 3. Results and discussion

### 3.1 Initial state of the c-Al$_{0.66}$Ti$_{0.33}$N coatings

A representative cross-sectional view of the as-coated arc-PVD c-Al$_{0.66}$Ti$_{0.33}$N nitrides on the Al$_2$O$_3$ substrates is displayed in Fig. 1. The average thickness of the nitride layer on Al$_2$O$_3$ was ~ 1.2 µm and the surface roughness of the c-Al$_{0.66}$Ti$_{0.33}$N-Al$_2$O$_3$ system was Ra = 1.84 µm ± 0.46. These surface characteristics are due to the intrinsic roughness of the polycrystalline alumina plate (Ra= 1.445 µm ±0.274) . The coating thicknesses were reached in a three-fold rotation configuration of the planetary system at 3 rpm. The Al/Ti average atomic ratio of the as-coated sample was about 2/1 matching the employed target composition (66/33). The coating surface showed no evidence of either micro-cracks or local delamination. XRD patterns of the as-coated c-Al$_{0.66}$Ti$_{0.33}$N on the Al$_2$O$_3$ plates displayed the diffraction peaks of the metastable cubic nitride phase. The peak reflexion at 63.3° 2θ corresponds to the (220) orientation of the cubic nitride phase. This peak and others close to this 2θ angle are commonly reported in the literature for similar coating compositions c-Ti$_{1-x}$Al$_x$N [21]. In our case, the XRD reflexion at 37.5° of the (111) plane of the coating is very close to the (110) plane of the substrate and the 43.45° 2θ corresponding to the (200) plane overlaps with the (113) plane of the alumina substrate (corundum JCPDS # 46-1212). The peak at 40° is related to the Ti-interlayer oriented in the (101) direction (Titanium JCPDS # 05-0682). No other crystalline phases such as c-TiN, c-AlN or h-AlN were detected for as-coating conditions.

XPS spectra of the as-coated c-Al$_{0.66}$Ti$_{0.33}$N system confirmed the presence of an Al-Ti-N bonds, see Fig. 2. The Al 2$p$ displays two unresolved peaks (Fig. 2a), the first peak found at 73.9 eV and the second at 75.2 eV indicating the existence of the Al-Ti-N bonds [22, 26] and Al-Ti-O-N bonds [26], respectively. No Al-N bonds in the range between 74.3 and 74.7 eV [27] were observed. Moreover, the Ti 2$p$ reveals mainly three species (Fig. 2b): the first doublet at 455.56 eV and 461.36 eV correspond to Ti 2$p_{3/2}$ and Ti 2$p_{1/2}$ branchings, respectively. This is in agreement with the binding energy of Ti in the TiAlN structure [22, 28], and no other signals of Ti-N bonds were found. The second doublet at 457.27 and 462.93 eV confirms the presence of TiO$_x$N$_y$ bonds [28] and, the third doublet at 459.20 and 464.86 eV is associated with Ti-O bonds into TiO$_2$ [29]. All the chemical information provided by the XPS spectra is in agreement with the crystal phases found by XRD. Fig. 2c displays the N 1s spectrum with three components at 396.59, 397.68 and 400.36 eV corresponding to Al-Ti-N, Al-Ti-O-N and N-C bonds [22, 26-28]. The last component at 400.36 eV has been related to either N-adsorbates [22, 26] or organic bonds N-C [26]. Furthermore, the O 1$s$ shows two unresolved peaks at 531.19 and 532.99 eV indicating the presence of the O-Ti and Al-O bonds, respectively [22-27, 30]. The presence of oxygen-containing species has been associated with at least two contributions: residual oxygen in the chamber (e.g. gases impurities, humidity, and adsorbed oxygen) that may react with the growing nitride [31] and the atmospheric oxygen present after deposition and cooling [32].

## 3.2 Thermal treatments of c-$Al_{0.66}Ti_{0.33}N$ in pure Ar.

Only a few TGA studies of c-$Ti_{1-x}Al_xN$ coatings have been reported analyzing their behavior under inert atmospheres and/or low-oxygen partial pressures. For instance, Martin Moser *et al* [33] and Rizzo *et al* [22] investigated the spinodal decomposition of c-AlTiN coatings using DSC and Differential Thermal Analysis (DTA) in inert atmospheres (i.e. 99.9 % helium and 99.999 % argon respectively) at a heating rate of 20 °C/min and no dwelling time. In those investigations, the AlTiN oxidation behavior was not the central point of the research and no data was reported in this regard. In our study, the arc-PVD c-$Al_{0.66}Ti_{0.33}N$ coatings were isothermally oxidized in 99.997 % pure Ar at 700, 800, 900, and 1000 °C during 5 h. At the selected heating rate of 50 °C/min, the set point temperature was reached within ~13 to 16 min. The normalized mass-gain curves after oxidation are compared in Fig. 3. A very similar parabolic mass-gain was observed upon treatments at 700 and 800 °C up to ~ 60 min (Fig. 3a). The linear trends observed after plotting mass gain vs. time$^{0.5}$ in Fig. 3b confirm their diffusion-controlled oxidation behavior (following a parabolic model) with a higher slope observed at 800 °C in comparison with 700 °C ($kp_{700}$ = 2.64333X10$^{-6}$ mg$^2$/min.cm$^4$) as a result of its faster oxidation rate ($kp_{800}$ = 4.34893X10$^{-6}$ mg$^2$/min.cm$^4$). In this stage, the inward diffusion of oxygen and outward transport of the Al and Ti species controls the oxidation process. The linear behavior observed at 700 and 800 °C suggest that the thermal growth oxide is mainly controlled by single phase oxide growth [33]. Furthermore, purely parabolic behavior with a $kp_{900}$ = 6.57865X10$^{-6}$ mg$^2$/min.cm$^4$) was recorded for the sample treated at 900 °C for times below 100 min, followed by a slight mass-gain decrease (Fig. 3b). The wavy nature of the TGA curve could be attributed to three main causes: the formation of gaseous nitrogen species being released, micro-spallation of tiny fractions of the porous oxide layer and signal-noise ratio from the TGA. On the other hand, the isothermal treatment at 1000 °C exhibited two linear oxidation kinetics (Fig. 3a), which indicates the non-protective nature of the oxide layer formed on the c-Ti1-xAlxN coatings. The first region, below 30 min, with a faster mass gain and $k_{l,t<30min}$ = 2.18774x10$^{-5}$ mg$^2$/min.cm$^4$ was followed by a mass gain decrease again indicating the non-protective nature of the oxide scale with $k_{l,t>30min}$ = 6.34517x10$^{-5}$ mg$^2$/min.cm$^4$. This oxidation behavior has been related to the absence of oxygen gradient in the coating [34]. In a similar way, the wavy behavior at 900 and 1000 °C could be also partially related to coating stress and micro-cracking followed by further oxidation reactions at the open surface cracks.

The spinodal decomposition of the c-Al$_{0.66}$Ti$_{0.33}$N is an additional phenomena that might affect the oxidation behavior (and mass gain) of the coating system at 800 °C and higher temperatures. The spinodal decomposition onset at ~ 800 °C depends on the coating composition [34]. Nitrides with high Al-content such as c-Al$_{0.66}$Ti$_{0.33}$N start to decompose into the cubic single nitrides at about 817 °C [19]. Consequent oxidation reactions of the single nitrides, c-TiN and c-AlN (at some point also h-AlN) of c-Al$_{0.66}$Ti$_{0.33}$N might be relevant above 900 °C considering the well-known susceptibility to oxidation of these binary nitrides.

In our study, the spinodal decomposition of the c-Al$_{0.66}$Ti$_{0.33}$N coating was not detected by the applied characterization techniques. However, the formation of the binary cubic nitrides and its effect on the oxidation cannot be fully ruled out and further investigations are needed to address these questions.

### *3.3 Phase evolution of the c-Al$_{0.66}$Ti$_{0.33}$N after Ar treatments.*

The evolution of the crystalline phases by XRD upon thermal treatment of the c-Al$_{0.66}$Ti$_{0.33}$N coating in pure Ar is shown in Fig. 4. The peak reflection at 63.54° corresponds to the (220) plane of the cubic ternary nitride (JCPDS # 00-080-4072), indicating that the c-Al$_{0.66}$Ti$_{0.33}$N phase is detected until 900 °C/5h in Ar whereas at 1000 °C/5h in Ar the peak shows a clear attenuation suggesting the nitride degradation, *e.g.*, spinodal decomposition and nitride oxidation. In the XRD data the peaks associated to α-Al$_2$O$_3$ (JCPDS # 01-730-5928) are mixed contributions coming from the substrate and the thermally grown oxide upon treatment. Initially, a decrease in the α-Al$_2$O$_3$ peaks are observed at 700°C followed by a continuous increase after 800°C. At 700 °C the cubic c-Al$_{0.66}$Ti$_{0.33}$N phase is still present and the 2θ peak = 34.6 ° appears and might be related to the cubic α-Al$_2$TiO$_5$ phase formation (JCPDS # 00-018-0068). This peak phase is better resolved after treatment at 800 °C suggesting further phase increase. Additionally, 2θ at ~ 37° may indicate oxidation processes contributing to evolve into c-Al$_{0.54}$Ti$_{2.46}$O$_{0.28}$N$_{4.58}$ crystals (JCPDS # 00-042-1279), which seems to develop up to 900 °C and then decreased at higher temperatures almost disappearing after the treatment at 1000 °C. At this temperature the peak at 2θ at 36.8° also points out to c-TiO phase. At 900 °C the coating exhibits the onset formation of rutile r-TiO$_2$ displaying the 2θ = 27.54°, 36.1°, and 54.26° diffraction peaks corresponding to the (110), (101) and (221) planes, respectively (JCPDS # 00-001-1292). The diffraction peaks related to the rutile r-TiO$_2$ phase are clearly identified after oxidation at 1000 °C/5h in Ar. These facts indicate that oxide phases such as AlTiO-N, α-Al$_2$TiO$_5$ and c-TiO could be related to the intermediate species that lead to the formation of thermodynamically stable rutile and aluminum oxides. In the literature, α-Al$_2$TiO$_5$ and c-TiO oxides were found after annealing of similar nitrides at 900 °C/5h [35, 36] and 1000 °C [37]. In summary, the XRD data display clear evidence of the Ti-based oxides being formed at 1000 °C. However, these data do not conclusively explain the oxide evolution at 900 °C and lower temperatures. The XPS data reported in next section provide better insight into the oxidation processes at those temperatures.

*XPS evaluation after the treatments in Ar*

The chemical composition of the thermally treated (oxidized) samples was analyzed by XPS. Surface XPS data for Al 2$p$, Ti 2$p$, N 1$s$ and O 1$s$ was acquired on the sample thermally treated at 700 °C, while depth profiling data up to ~ 500 nm was recorded for samples treated at 800 °C and 900 °C. The XPS analysis of the sample thermally treated at 700 °C/5h in Ar presented slight oxidation of c-Al$_{0.66}$Ti$_{0.33}$N coating. Well-defined O 1$s$ signals and no BE peaks due to N 1$s$ or Ti 2$p$ were found showing the N-depletion and the formation of only aluminum oxide on the coating surface. This is in agreement with the observations made by McIntyre *et al* [38] and the theory of mass loss caused by N-release.

*XPS depth profiling*

The XPS depth profiles were acquired for the samples treated at 800 and 900 °C. At these temperatures no clear indication of coating oxidation was detected by the low-angle XRD analysis. Figures 5 to 8 present the XPS depth profiling analysis for the c-Al$_{0.66}$Ti$_{0.33}$N coatings after TGA. BE of Al 2$p$ recorded at 800 and 900 °C display the thermally grown oxide evolution along the coating thickness in Fig. 5. Al 2$p$ signals at ~ 74.4 eV below ~ 92.5 nm were associated with γ−Al$_2$O$_3$ according to Rotole and Sherwood [39]. Above ~ 246.8 nm of depth, and additional small peak was identified for both temperatures at higher energies, ~ 74.84 eV. This peak was indexed as α-Al$_2$O$_3$ according to the theoretical work reported by Lizarraga et al [40], where XPS spectra were calculated for amorphous alumina and its polymorphic phases based on first-principles calculations. In this study, a theoretical BE shift of 0.54 eV was reported from the single α-Al$_2$O$_3$ peak corresponding to the six fold aluminum coordination number Al(VI) to the central fourfold aluminum coordination number Al(IV) of the γ−Al$_2$O$_3$ peak. This ΔeV is in good agreement with the experimental shift observed in our measurements; nevertheless, variations can be expected either by the deviations for the stoichiometry and the XPS experimental setup. For instance, the γ−Al$_2$O$_3$ phase can admit nitrogen in solution and still maintain its cubic structure, e.g., Al$_{2.85}$O$_{3.45}$N$_{0.55}$ JCPDS # 01-080-2171. This phase could not be confirmed by XRD in Fig. 4 due to the overlap of its main peaks (311) and (440) with the α−Al$_2$O$_3$ phase signal coming from the substrate and the new α−Al$_2$O$_3$ formation from the coating as a result of the thermal treatments. The confirmation of γ−Al$_2$O$_3$ formation by XPS, in combination with the parabolic oxidation behavior observed for the samples treated at 800, $k_{p_{800°C}}$ = 4.349 x 10$^{-6}$ mg$^2$/cm$^4$ min below 5h, and 900 °C, $k_{p_{900°C}}$ = 6.578 x 10$^{-6}$ mg$^2$/cm$^4$min below 100 min, suggest that the thermally grown oxide kinetics are mainly controlled by the growth of γ−Al$_2$O$_3$ in the parabolic region. This, in turn, is supported by the parabolic rate constants values reported by Brumm and Grabke in [41] for γ−Al$_2$O$_3$ growth, $k_{p_{800°C}}$ ~ 1.4 x 10$^{-6}$ mg$^2$/cm$^4$min and $k_{p_{900°C}}$ ~ 6 x 10$^{-6}$ mg$^2$/cm$^4$min . At the same time, an increase in the peak signal of Al-Ti-O-N compounds is observed as a function of the depth profile, which was associated with the c-Al$_{0.66}$Ti$_{0.33}$N coating. An oxygen enrichment of the c-Al$_{0.66}$Ti$_{0.33}$N is expected upon thermal treatments and even the formation of transient oxy-nitrides compounds.

Similarly, O 1*s* peaks were also in good agreement with the expected BE of ~ 531.8 eV for γ−Al$_2$O$_3$ [39] after treatments at 800 and 900°C, as shown in Fig. 6. In our study, the O 1*s* peaks displayed slight chemical shifts at the coating depth that might be associated to oxygen bonded to Al-Ti-N-containing species which agrees with the XRD results showed in Fig. 4 where the Al$_{0.54}$ Ti$_{2.46}$ N$_{0.28}$ O$_{4.58}$ phase (JCPDS # 00-042-1279) was indexed at 800 and 900°C.

Ti 2p peaks at ~ 61 and 92 nm were extremely weak and could not be resolved at 800 °C. This result suggests that Ti-containing species below 92.5 nm of depth are in the range of the XPS detection limit, i.e., from 0.1 to 1 at. %, and therefore, that a protective Al$_2$O$_3$ scale is still present after 5 h thermal treatment. This assumption was confirmed with the analysis of Fig 9. Fig. 7a shows oxidized regions located between 246 and 500 nm contain Ti-species associated to Ti-O (463.95 and 458.35 eV), Ti-Al-O-N (462.54 and 456.75 eV) and Ti-Al-N (460.75 and 455.14 eV) bonds. On the other hand, Fig. 7b displays the Ti 2*p* spectra of the Ti-species formed upon 5 h oxidation treatment at 900 °C. BE at 455 eV in the first 61.5 nm is due to the presence of Al-Ti-N bonds. A doublet at 462.88 and 457.28 eV associated with Ti-Al-O-N species was also observed with a slightly shift to higher binding energies suggesting a chemical change relative to Ti-species at deeper regions. At ~ 92.5 nm the signals of the Ti-Al-O-N bonds were 462.2 and 456.6 eV and remain constant at coating depths up to 500 nm. BEs at 462.2 and 456.6 eV have been associated to sub-stoichiometric TiOx oxides [26] or to $TiO_xN_y$ bonds [25, 28]. A similar trend of the BEs was revealed for Ti 2*p* corresponding to Ti-O species. BEs of Ti-O were ~ 459 and 464.68 eV near the surface of the oxide scale (~ 61.7 nm), and slight shift to higher BEs was detected. At ~ 92.5 nm, the doublet of Ti-O bonds was found at 458 and 463.6 eV, which remain unchanged at the coating depth up to 500 nm. This analysis evidences that a non-continuous but homogenous alumina scale is obtained after 5 h oxidation at 900 °C. On the contrary, various oxides and oxy-nitrides such as Ti-rich oxides ($TiO_2$, $TiAlO_x$, Ti-Al-O-N) and Al-rich oxides ($Al_2O_3$, $AlTiO_x$, AlTiON) may co-exist up to 500 nm after treatment at 900 °C. According to the XPS spectra of Al- and Ti-containing species, gradual transitions among oxides inside the scale may be expected.

N 1*s* peaks after depth profiling of the oxidized coatings treated at 800° and 900 °C are shown in Fig. 8. Similarly to the Ti 2*p* signals, N 1s peaks were not observed at ~ 62.7 and 92.5 nm of depth after treatment at 800 °C, and only at ~ 246.8 nm of depth, Al-Ti-N and Al-Ti-N-O bonds were observed at ~ 397 eV [22, 26, 29]. At further depths, the intensity of the N 1s peak increased and starting from the near surface scale these observations represented the N-depletion in the coating, see Fig. 8a. Less significant N-depletion was detected on the coating treated at 900

°C, where BE of N 1*s* assigned to the Al-Ti-N becomes visible already at ~ 64 nm and along the whole sputtered scale thickness in the coating, Fig. 8b.

In summary, a less protective oxide scale is observed for the coating treated a 900 °C, where a mixed- thermally grown oxide scale was obtained after 5 h of treatment. This can be the result of the lack of Al-flux to sustain the continuous growth of the $Al_2O_3$ scale and the simultaneous thermal decomposition of the $Al_{0.66}Ti_{0.33}N$ coatings into its single nitrides and with subsequent $N_2$-release. The latter could be an explanation for mass-gain loss after 225 min of oxidation observed in Fig. 3, in addition to the local spallation events.

*TEM evaluation after treatment in Ar at 1000 °C*

Even though the c-$Al_{0.66}Ti_{0.33}N$ coating withstands the thermal treatment in Ar at 1000 °C/5h, the partial oxidation of the nitride coating was evident after the TGA, XRD and XPS analysis. Fig. 9 shows the cross-sectional view of the sample treated in Ar at 1000 °C/5h, where four different oxide layers can be clearly observed along the thermally grown oxide thickness. STEM images show a ~ 90-nm aluminum oxide layer formed at the top coating (zone V), only Ti-traces (less than 0.5 at. %) were measured in this zone. Beneath the $Al_2O_3$ layer, a porous Al-rich layer of ~ 155 nm was found (zone IV). In this zone, the layer is based on Al-rich oxides with relatively low Ti contents (< 6 at. %) and clear absence of nitrogen. The porous structure of this layer can be related with the N2-release as a result of the c-$Al_{0.66}Ti_{0.33}N$ coating degradation upon treatment. The next oxide layer (zone III) appeared at the coating depth of about 245 nm with a compact (Al,Ti)-rich oxide layer structure, but with considerably higher Ti-content and N-traces. At ~ 330 nm, a Ti-rich layer with low Al-content was identified (zone II). This layer holds high amounts of nitrogen and about half oxygen content compared to the first three oxide layers near the top surface. Table I summarizes the average chemical composition in at. % of each oxidized layer of the nitride coating after treatment at 1000 °C/5h in Ar. It is important to consider that the STEM image has roughly estimated oxides and oxi-nitrides stoichiometry based on the average chemical composition of the TEM-EDS analysis.

Table II displays various $Al_xTi_{1-x}N$ (x = 0.47, 0.5, 0.6, 0.66, 0.7) nitride compositions and thermal treatments conditions (gas, temperatures and times). Most of the oxidation/treatment investigations have been carried out in either pure oxygen or air and only Yang et al [21] reported oxidation results after treatment in Ar. This analysis displays that mainly α-$Al_2O_3$ and rutile and anatase $TiO_2$ phase(s) are formed upon oxidation in air, this is due to the high oxygen content, and only few investigations have suggested the formation of complex $Al_xTi_yO_z$ oxides [21] under low-oxygen atmospheres. No further information about the oxides and transition phases and/or compositions was mentioned. In addition, none of them reported the presence of oxy-nitrides phases. Our study suggests that oxidation under low oxygen content follows a complex reaction pathway, probably forming intermediate species such as oxy-nitrides (Al-rich, Al + Ti-rich, Ti-rich) and oxides (i.e. AlTiON, AlTiO, $Al_2TiO_5$) before full oxidation into α-$Al_2O_3$ (containing Ti traces) and rutile $TiO_2$.

*3.4 Pin on disk tests in Ar jet of the arc PVD c-$Al_{0.66}Ti_{0.33}$N-AISI M2 system*

Friction coefficient vs. distance of the arc PVD $Al_{0.66}Ti_{0.33}N$ coating as a function of temperature in Argon jet is shown in Fig. 10. Considering the tribological pair (α-$Al_2O_3$ ball and c-$Al_{0.66}Ti_{0.33}N$ coating), their poisons ratio (ν), E-module (E) and the applied load of 5 N, an estimated Hertzian contact pressure of 1.64 GPa was reached. At room temperature, a stable CoF = 0.2 up to 10,000 cycles was obtained. A relative short stage of asperities removal was observed during this test at RT and relatively low CoF values were obtained through the pin-on-disk test in Ar. Optical images of the coatings (Fig 11a.) indicate adhesion in the center of the wear track due to plastically deformed particles that cause ploughing of the surface. Cracks result from abrasive wear due to the presence of particles with high hardness and induced stresses during ploughing. Crack opening, oriented opposite to the pin-on-disk movement indicate that cracks are caused by tensile stresses. Upon heating at 700 and 800 °C in Ar, $Al_{0.66}Ti_{0.33}N$ reached a CoF ~ 0.6 for about 50 m with short stages of asperity removal; then, CoF decreased to 0.2 at 700 °C and 0.35 at 800 °C after 10,000 cycles. Instabilities and local variations on the CoF are associated to the dynamical nature of the wear test. CoF decrease is related to particles being formed and removed inside the wear track where wear behavior might be the result of brittle and/or hard oxide layer(s) removal, while nitride and oxy-nitride particles present high hardness and certain plastic deformation resistance.

Oxide crystals grown upon temperature and time may act as lubricant layers and might be partially or fully removed at some point of the wear track resulting in CoF changes. Phases such as $TiO_2$ and sub-stoichiometric $TiO_x$ oxides are related to CoF decrease and increased lubricity [41]. The test performed at 700 °C (Fig. 11b) shows a reduction of particle adhesion on the surface, these are signs of micro-scratching and the main reason of cracking. Micro-plough identified in our tests is associated to wear caused by relatively small (generally micro-size order) hard particles, while cracking randomly oriented indicates fatigue and embrittlement of the substrate due to the high temperature. The wear track of the coating tested at 800 °C (Fig. 11c) showed a morphology similar to that observed at 700 °C, cracks are randomly oriented and might be related to steel substrate fatigue and embrittlement. However, feathery like cracks oriented opposite to sample rotation are observed at the track edges. These type of cracks are produced by tensile stresses induced in the back of the pin during rotation on the surface. Moreover, elasticity reduction of the system due to applied high temperature must be taken into account. Additionally oxidation occurred to some extent, and it is possible that the $Al_{0.66}Ti_{0.33}N$ coating surface becomes fragile promoting this failure mode. The coating worn at 900 °C (Fig. 11d) showed a rarefied surface in the wear track with a large number of small particles, presumably oxides adhered to the surface formed by the high temperature and tribo-corrosion, which produced coating micro-scratching and ploughing. No cracking on the coating surface, associated with fatigue, reduced elasticity and increased fragility were observed. This is an indication that particle adhesion might be responsible for the stable CoF values (0.55 – 0.6) at 900 °C. Wear track inspection suggested high wear and abrasion as the predominant processes resulting from relatively high contact stresses applied on the surface (1.642 GPa). On the other hand, although the coating was tribo-protected with argon jet, the increased working temperature promoted oxidation and elasticity decrease of the coating-substrate system. These two phenomena are associated with brittle failure mode and surface fatigue. At 900 °C, a large amount of small coating particles adhered to the wear track, oxide and oxy-nitrides mixtures, maintained the CoF below 0.6.

*3.5 XRD evaluation after Tribology in Ar-jet*

Grazing incidence XRD analysis of the $Al_{0.66}Ti_{0.33}N$-AISI M2 steel samples after pin-on-disk tests are presented in Fig. 12. After test at RT, peaks of the c-$Al_{0.66}Ti_{0.33}N$ coating are clearly identified and no indications of oxidation processes were found. Surface amorphization after samples tests at 700 and 800 °C could be related to precursor phases before forming stable oxides, see $2\theta$ values between ~ 50 and 60° in Fig. 12. First indications of $Al_2O_3$ formation ($2\theta$ = 25.3°) and $TiO_2$ ($2\theta$ = 54.3°) are detected after tribology at 800 °C. These oxides evolve and growth clearly after testing at 900 °C.

*3.6 Scratch tests after Tribology in Ar-jet*

Critical loads of the c-Al$_{0.66}$Ti$_{0.33}$N coatings after tribology in Ar jet are shown in Fig. 13. At RT the coating displayed cohesive failures composed of conforming cracks, typical of a hard coating deposited on a ductile substrate, Fig. 14a. Upon load increase, recovery spallation at the edges of the scratch track is observed. This is caused by the substrate elastic recovery and cohesive cracking of the coating. In the scratch center, signs of buckling which are characterized by irregularly spaced arc-like patches and partial coating delamination appeared. These defects are due to fragile characteristics of the nitride; however, no complete delamination of the coating was observed, Fig. 14. This result indicates that regardless the high contact pressure applied during scratch test, the coating-substrate system has beneficial synergy in terms of adhesion strength. Upon temperature increases, progressive reduction of cohesive critical loads (Lc1) were observed. This effect is characterized by buckling-type spallation of the coating, see Fig. 13. On the other hand, critical adhesive loads (Lc2) composed mainly of wedging-type spallation were observed in the scratch track, Fig. 14. Detrimental effects on adhesion and brittleness raise occur due to the localized formation of oxides which were detected by XRD analysis, which produce composition gradients, increasing surface tensions and internal stress in the coating. In addition, residual stress increases generated by the thermal expansion of the system, tend to affect interfacial coherence and reduce the fracture toughness [42, 43]. However, in the present adhesion tests full delamination of the oxidized coatings in argon jet was not observed. This indicates that despite the fragile nature of the oxide scale and the adhesive faults measured during scratch, the coating was still strongly bonded to the steel.

## 4. Conclusions

The c-Al$_{0.66}$Ti$_{0.33}$N coatings were deposited on polycrystalline alumina plates by applying the arc-PVD method. In the as-coated state, the c-Al$_{0.66}$Ti$_{0.33}$N layer already contained a small fraction of oxygen. These coatings withstand the exposure to high purity Ar (99.997 %) at 700 and 800 °C, showing almost no oxidation, only detected by the XPS analysis. In general, the TGA curves under these experimental conditions are very sensitive to changes upon the temperature treatment and surely affected by phenomena such as micro-crack formation and partial coating micro-spallation due to the alumina surface roughness. The linear oxidation behavior observed at 700 and 800 °C, suggest that oxidation reaction is mainly controlled by single phase oxide growth. Parabolic mass-gain was measured at temperatures below 900 °C, while linear mass-gain was observed at 1000 °C suggesting the oxidation process not being limited by diffusion.

The obtained results showed that using Ar (low oxygen content) can slow down the reactions occurring with oxygen and shed light into the details and paths of the nitride oxidation processes. According to the XRD-evaluation, the nitrides were only slightly oxidized at 900 °C and more severely oxidized at the 1000 °C/5h treatment. XRD diffraction inspection between 700 and 900 °C indicate that oxide species such as c-TiO, c-$Al_{0.54}Ti_{2.46}O_{0.28}N_{4.58}$ and α-$Al_2TiO_5$ might participate during coating oxidation; and at certain temperatures (~ 850° - 900 °C) the *spinodal decomposition* of the nitride coating must be considered as well. Different oxidation scenarios are possible considering the *non-spinodal region* (at 800 °C and below) where nitrides might oxidize without phase decomposition and the *spinodal region* (at 900 °C and higher) where c-$Al_{0.66}Ti_{0.33}N$ first decomposes into c-AlN (or h-AlN) + c-TiN and then these phases might oxidize.

The XPS analysis of the depth profiling of the oxidized c-$Al_{0.66}Ti_{0.33}N$ coatings provided additional insights into the oxides being formed. According to the detected BEs, the coating surface was already slightly oxidized at 700 °C/5h in Ar. At this temperature, the nitride transformed into an Al-based oxide and no other oxides were detected at the top surface. But, an open question is, if there is any α-alumina being formed at this temperature and time.

The depth profiling analysis of the coatings exposed to Ar at 800 and 900 °C indicate the co-existence of both Ti- and Al-based oxides such as $TiO_2$, $AlTiO_xN_y$, $\alpha-Al_2O_3$ and $\gamma-Al_2O_3$ in the scale. The c-$Al_{0.66}Ti_{0.33}N$ nitride partially withstood the treatments in Ar up to 900 °C. At this temperature, the nitride transformed into the rutile and probably an additional AlTiON oxy-nitride phase with low Ti-content. Interestingly, the oxide layers displayed no sharp boundary edges, but gradual transitions from one oxide to the next phase starting with an $Al_2O_3$-rich zone near the top surface, followed-by a mix-oxide region and lastly, a $TiO_2$-rich zone at 500 nm coating depth. The thermal treatment at 1000 °C/5h resulted in the oxidation of about one third of the total coating thickness. This sample clearly showed four oxide and oxy-nitride layers; $Al_2O_3$ at the top (probably α), succeeded by an Al-rich oxide layer with low Ti-content (< 6 at. %), then an (Al, Ti)-rich oxy-nitride (~ 32.4 at. % Al) containing N-traces (< 3 at. % N) and a transition (Ti, Al)-rich oxy-nitride layer with low Al-content (~ 7 at. % Al) at the boundary with the coating nitride. These findings display the complexity of the oxidation processes of c-$Al_{0.66}Ti_{0.33}N$ at the microscopic level.

Friction coefficient increased upon temperature increments showing relatively short stages of asperity removal; however, particle abrasion dominated the wear behavior during tribology tests. At RT, combined effects of high scratching, ploughing and adhesion due to hard particles with low plastic deformation was observed. As the temperature increased, a mixture of small particle brittle oxides, hard nitrides and oxi-nitrides are formed, which induced micro-ploughing, abrasion and adhesion damage. High temperature pin-on-disk tests showed high oxide/oxi-nitride adhesion on the coating surface preventing friction coefficient to further diminish during the stabilization stages.

Adhesion tests showed a slight reduction of cohesive and adhesive critical loads after exposure to temperature and Ar. These changes are associated to coating degradation near the top surface due to fragile top oxide layers (p.ej. rutile at 900 °C) and formation of various Al-rich and Ti-rich oxides/oxy-nitrides layers. At some stage, scratch load increase and penetrated into the coating upon load evolution, displaying the characteristic failure modes such as buckling and wedge spallation of the hard nitride deposited on a plastic substrate. The absence of gross spallation, or complete separation of the coating, that could be described as Lc3, indicate that despite the identified fragile failures, the coating-substrate still preserved high synergy after exposure to temperature in Ar.

Acknowledgments:

This work was supported by National Council of Science and Technology (CONACyT), through the program "Frontiers of Science" and the project No: 2015-02-1077, and the program "Cátedras Conacyt".

The authors thank the National Lab – CENAPROT for providing all the facilities and use of the arc-PVD coater unit at CIDESI. The authors also would like to thank the National Laboratory of Nanotechnology (NaNoTeCh) from CIMAV for the TEM analysis.

The X-ray photoelectron spectra were obtained by Dra. Mariela Bravo-Sanchez at the National Laboratory of Research in Nanoscience's and Nanotechnology (LINAN) at IPICYT, S.L.P., México.

Thanks to Ing. Carolina Ortega Portillo & Maria José Gaytan Sánchez for performing the pin-on-disk tests.

References:

[1] P.H. Mayrhofer, L. Hultman, J.M. Schneider, P. Staron, H. Clemens, Spinodal decomposition, of cubic Ti$_{1-x}$Al$_x$N: Comparison between experiments and modeling. Int. Mat. Res. 98 (2007) 11, 1054-1059.

[2] T. Ikeda & H. Satoh, Phase formation and characterization of hard coatings in the Ti-Al-N system prepared by the cathodic arc ion plating method, Thin Solid Films, 195 (1991) 99-110.

[3] L. Hultman, Thermal Stability of nitride thin films, Vacuum 57 (2000) 1-30.

[4] P.H. Mayrhofer, A. Hörling, L. Karlsson, J. Sjölén, T. Larsson, C. Mitterer, L. Hultman, Self-organized nanostructures in the Ti–Al–N system, Appl. Phys. Lett. 83 (2003) 2049-2051.

[5] P.H. Mayrhofer, F.D. Fischer, H.J. Böhm, C. Mitterer, J.M. Schneider, Energetic balance and kinetics for the decomposition of supersaturated Ti$_{1-x}$Al$_x$N, Acta Materialia 55 (2007) 1441-1446.

[6] N. Norrby, Microstructural evolution of TiAlN hard coatings at elevated pressures and temperatures, PhD-Thesis, Linköping University (2011) 21-24.

[7] J. Zhou, J. Zhoung, L. Chen, L. Zhang, Y. Du, Z. Liu, P.H. Mayrhofer, Phase equilibria, thermodynamics and microstructure simulation of metastable spinodal decomposition in c-Ti$_{1-x}$Al$_x$N coatings, CALPHAD: Computer Coupling of Phase Diagrams and Thermochemistry 56 (2017) 91-101.

[8] A. Escudeiro Santana, A. Karimi, V.H. Derflinger, A. Schütze, The role of hcp-AlN on hardness behaviour of Ti$_{1-x}$Al$_x$N nanocomposite coatings, Thin Solid Films 469-470 (2004) 339-344.

[9] N. Norrby, L. Rogström, M.P. Johansson-Joesaar, N. Schnell, M. Odén, In situ X-ray scattering study of the cubic to hexagonal transformation of AlN in Ti$_{1-x}$Al$_x$N. Acta Materialia 73 (2014 2015-214.

[10] R. Rachbauer, S. Massl, E. Stergar, D. Holec, D. Kiener, J. Keckes, J. Patscheider, M. Stiefel, H. Leitner & P.H. Mayrhofer, Decomposition pathways in age hardening of TiAlN films, J. Appl. Phys. 110, (2011) 1-10.

[11] Y.H. Yoo, D.P. Le, J.G. Kim, S.K. Kim, P.V. Vinh, Corrosion behaviour of TiN, TiAlN, TiAlSiN thin films deposited on tool steel in the 3.5wt.% NaCl solution, Thin Solid Film 516 (2008) 3544-3548.

[12] D.M. Devia, E. Restrepo-Parra, P.J. Arango, A.P. Tschiptschin, J.M. Velez, TiAlN coatings deposited by triode magnetron sputtering varying the bias voltage, Appl. Surf. Sci. 257 (2011) 6181-6185.

[13] N.D. Nam, M. Vaka, N.T. Hung, Corrosion behaviour of TiN, TiAlN, TiAlSiN-coated 316L stainless steel in simulated proton exchange membrane fuel cell environment, J. Power Sources 268 (2014) 240-245.

[14] H. Elmkhah, T.F. Zhang, A. Abdollah-zadeh, K.H. Kim, F. Mahboudi, Surface characteristics for the Ti-Al-N coatings deposited by high power impulse magnetron sputtering technique at the different bias voltages, J. Alloys Compd. 688 (2016) 820-827.

[15] W.D. Münz, Titanium aluminum nitride films: A new alternative to TiN coatings, J. Vac. Sci. & Technol. A 4, (1986) 2717-2725.

[16] M. Kawate, A.K. Hashimoto, T. Suzuki, Oxidation resistance of $Cr_{1-x}Al_xN$ and $Ti_{1-x}Al_xN$ films, Surf. & Coat. Technol., 165 (2003) 163-167.

[17] P.C. Jindal, A.T. Santhanam, U. Schleinkofer, A.F. Shuster, Performance of PVD TiN, TiCN, TiAlN coated cemented carbide tools in turning, Int. J. Refract. Met. H. 17 (1999) 163-170.

[18] J. Vetter, Vacuum arc coatings for tools: potential and application, Surf. & Coat. Technol. 76-77 (1995) 719-724.

[19] A. Knutsson, I.C. Scharmm, K.A. Groenhagen, F. Muecklich and M Odén, Surface directed spinodal decomposition of TiAlN/TiN interfaces, J. Appl. Phys. 113 (2013) 114305.

[20] M.P. Johansson Joesaar, N. Norrby, J. Ullbrand, R. M´Saoubi, M. Odén, Anisotropy effects on microstructure and properties in decomposed arc evaporated $Ti_{1-x}Al_xN$ coatings during metal cutting, Surf. & Coat. Technol. 235 (2013) 181-185.

[21] B. Yang, L. Chen, K.K. Chang, W. Pan, Y.B. Peng, Y. Du, Y. Liu, Thermal and thermos-mechanical properties of TiAlN and CrAlN coatings, Int. Journal of Refractory Metals and Hard Materials. 35 (2012) 235-240.

[22] A. Rizzo, L. Mirenghi, M. Massaro, U. Galietti, L. Capodieci, R. Terzi, L. Tapfer, D. Valerini, Improved properties of TiAlN coatings through the multilayer structure, Surf. & Coat. Technol. 235 (2013) 475-483.

[23] U.S. Dixit, D.K. Sarma, J. Paulo Davim, Environmentally Friendly Machining, Springer.

[24] E.O. Ezugwu, R.B. Da Silva, J. Bonney, A.R. Machado, The Effect of Argon-Enriched Environment in High-Speed Machining of Titanium Alloy, Tribology Transactions, 48: (2005) 18-23.

[25] F.F. Lima, W.F. Sales, E.S. Costa, F.J. da Silva, A.A.R. Machado, Wear of ceramic tools when machining Inconel 751 using argon and oxygen as lubri-cooling atmospheres, Ceramics International 2017.

[26] Z.T. Wu, P. Sun, Z.B. Qi, B.B. Wei, Z.C. Wang, High temperature oxidation behaviour and wear resistance of $Ti_{0.53}Al_{0.47}N$ coating by cathodic arc evaporation, Vacuum 135 (2017) 34-43.

[27] NIST X-ray Photoelectron Spectroscopy Database, https://srdata.nist.gov/xps.

[28] G. Greczynski, J. Jensen, J.E. Green, I. Petrov, L. Hultman, X-ray Photoelectron Spectroscopy Analyses of the Electronic Structure of Polycrystalline $Ti_{1-x}Al_xN$ Thin Films with $0 \leq x \leq 0.96$ Surf. Sci. Spec., Vol. 21, (2014) 35-49.

[29] M.C. Biesinger, L.W.M. Lau, A.R. Gerson, R.St.C. Smart, Resolving surface chemical states in XPS analysis of first row transition metals, oxides and hydroxides: Sc, Ti, V, Cu and Zn, Appl. Surf. Sci. 257 (2010) 87-898.

[30] Z.B. Qi, P. Sun, F.P. Zhu, Z.T. Wu, B. Liu, Z.C. Wang, D.L. Peng, C.H. Wu, Relationship between tribological properties and oxidation behaviour of $Ti_{0.34}Al_{0.66}N$ coatings at elevated temperature up to 900°C, Surf. & Coat. Technol. 231 (2013) 267-272.

[31] C. Gnoth, C. Kunze, M. Hans, M. Baben, J. Emmerlich, J.M. Schneider and G. Grundmeier, Surface Chemistry of TiAlN and TiAlNO coatings deposited by means of high power pulsed magnetron sputtering, J. Appl. Phys. 46 (2013) 1-7.

[32] M. Braic, V. Braic, M. Balanceanu, G. Pavelescu, A. Vladescu, I. Tudor, A. Popescu, Z. Borsos, C. Logofatu, C.C. Negrila, Microchemical and mechanical characteristics of arc plasma deposited TiAlN and TiN/TiAlN coatings, J. of Optoelectronics and Adv. Mats. 7 (2005) 671-676.

[33] M. Moser, D. Kiener, C. Scheu, P.H. Mayrhofer, Influence of Yttrium on the Thermal Stability of Ti-Al-N Thin Films, Materials 3 (2010) 1573-1592.

[34] B. Xiao, H. Li, H. Mei, W. Dai, F. Zuo, Z. Wu, Q. Wang, A study of oxidation behaviour of AlTiN- and AlCrN-based multilayer coatings, Surf. & Coat. Technol. 333 (2018) 229 237.

[35] Y.X. Xu, L. Chen, B- Yang, Y.B. Peng, Y. Du, J.C. Feng, F. Pei, Effect of CrN addition on the structure, mechanical and thermal properties of Ti-Al-N coating, Surf. & Coat. Technol. 235 (2013) 506-512.

[36] M. Pfeiler, J. Zechner, M. Penoy, C. Michotte, C. Mitterer, M. Kathrein, Improved oxidation resistance of TiAlN coatings by doping with Si or B, Surf. & Coat. Technol. 203 (2009) 3104-3110.

[37] M. Pfeiler, C. Scheu, H. Hutter, J. Schnöller, C. Michotte, C. Mitterer, M. Kathrein, On the effect of Ta on improved oxidation resistance of Ti-Al-Ta-N coatings, J. Vac. Sci. Tech. A 27 (2009) 554-560.

[38] D. McIntyre, J.E. Greene, G. Håkansson, J.E. Sundgren, W.D. Münz, Oxidation of metastable single-phase polycrystalline $Ti_{0.5}Al_{0.5}N$ films: Kinetics and mechanisms, J. Appl. Phys. 67 (1990) 1542-1553.

[39] J. A. Rotole & P. M.A. Sherwood, Gamma-Alumina ($\gamma$-$Al_2O_3$) by XPS, Surf. Sci. Spectra 5, 18 (1998) 18–24.

[40] J. A. Rotole & P. M.A. Sherwood, Corundum ($\alpha$-$Al_2O_3$) by XPS, Surf. Sci. Spectra 5, 11 (1998) 11–17.

[41] T. Polcar, T. Kubart, R. Novak, L. Kopecký, P. Siroký, Comparison of tribological behavior of TiN, TiCN and CrN at elevated temperatures, Surf. Coat. & Technol. 193 (2005) 192-199.

[42] C. Badini, S. M. Deambrosis, E. Padovano, M. Fabrizio, O. Ostrovskaya, E. Miorin, G. C. D'Amico, F. Montagner, S. Biamino and V. Zin, Thermal Shock and Oxidation Behavior of

HiPIMS TiAlN Coatings Grown on Ti-48Al-2Cr-2Nb Intermetallic Alloy. Mats 9, 961 (2016) 1-20.

[43] A.E. Santana, A. Karimia, V.H. Derflinger, A. Schutze, Thermal treatment effects on microstructure and mechanical properties of TiAlN thin films. Tribol. Lett. 17, (2004) 689–696.

List of figure captions

Figure 1. STEM micrographs of the c-$Al_{0.66}Ti_{0.33}N$ on $Al_2O_3$ plates (as coated).

Figure 2. XPS spectra for (a) Al 2*p*, (b) Ti 2*p*, (c) N 1*s* & (d) O 1*s* for c-$Al_{0.66}Ti_{0.33}N$ (as coated).

Figure 3. Isothermal TGA´s of the c-$Al_{0.66}Ti_{0.33}N$ arc-PVD coatings in pure Ar/5h at each treatment temperature.

Figure 4. XRD patterns for the c-$Al_{0.66}Ti_{0.33}N$ coating on $Al_2O_3$ plates after TGA experiments in pure Ar at different temperatures. In this data the α-$Al_2O_3$ comes from the substrate and the oxidized coating.

Figure 5. Al 2*p* spectra after 5 h treatment in Ar at 800 °C (a) and 900 °C (b) as a function of coating depth.

Figure 6. O 1*s* spectra after 5 h treatment in Ar at 800 °C (a) and 900 °C (b) as a function of coating depth.

Figure 7. Ti 2*p* spectra after 5 h treatment in Ar at 800 °C (a) and 900 °C (b) as a function of coating depth.

Figure 8. N 1*s* spectra after 5 h treatment in Ar at 800 °C (a) and 900 °C (b) as a function of coating depth.

Figure 9. STEM images of the isothermally treated sample at 1000 °C/5h in Ar.

Figure 10. Friction coefficient (CoF) of arc PVD c-$Al_{0.66}Ti_{0.33}N$ coating on AISI M2 steel as a function of wear temperature in argon jet.

Figure. 11. Wear tracks of $Al_{0.66}Ti_{0.33}N$ coatings after tribology test in Ar jet at RT (a), 700 (b), 800 (c) and 900 °C (d). Optical images taken at 5 X and the inserts at 20 X.

Figure. 12. Grazing incidence XRD of arc PVD $Al_{0.66}Ti_{0.33}N$-AISI M2 system after pin on disk tests in Ar jet vs. working temperature.

Figure. 13. Critical loads of arc PVD c-$Al_{0.66}Ti_{0.33}N$-AISI M2 system after tribology in argon jet as a function of temperature.

Figure. 14. Scratch tracks on arc PVD c-$Al_{0.66}Ti_{0.33}N$-AISI M2 system after tribology in argon jet at (a) RT, (b) 700°, (c) 800° and (d) 900 °C.

List of tables

Table I. Chemical composition by EDS-TEM in at. % of $Al_{0.66}Ti_{0.33}N$ after treatment at 1000 °C/5h in Ar. The zone numbers correspond to the areas labeled in figure 8.

Table II. Effect of temperature, time and gas exposure on the oxidation behavior of c-$Ti_xAl_{1-x}N$ coatings, data published in the literature vs. present work.